\documentclass[twocolumn,showpacs,preprintnumbers,amsmath,amssymb,prb,longbibliography]{revtex4-1}

\usepackage{ucs}
\usepackage[utf8x]{inputenc}
\usepackage{amsmath}
\usepackage{amssymb}
\usepackage{bm}
\usepackage{fontenc}
\usepackage[dvips]{graphicx}
\usepackage[caption=false]{subfig}

\usepackage{color}

\newcommand{\expec}[1]{\langle #1 \rangle}

\begin{document}

\title{Dynamic RKKY interaction between magnetic moments in graphene nanoribbons}

\author{F. S. M. Guimar\~aes$^{1}$, J. Duffy$^{2}$, A. T. Costa$^{3}$, R. B. Muniz$^{3}$, M. S. Ferreira$^{2}$}
\affiliation{$^1$ Peter Gr\"unberg Institut and Institute for Advanced Simulation, Forschungszentrum J\"ulich \& JARA, D-52425 J\"ulich, Germany\\ 
$^2$ School of Physics, Trinity College Dublin, Dublin 2, Ireland\\ 
$^3$ Instituto de F\'{\i}sica, Universidade Federal Fluminense, Niter\'oi, Brazil}

\date{\today}

\begin{abstract}

    Graphene has been identified as a promising material with numerous applications, particularly in spintronics. 
    In this paper we investigate the peculiar features of spin excitations of magnetic units deposited on graphene nanoribbons and how they can couple through a dynamical interaction mediated by spin currents. 
    We examine in detail the spin lifetimes and identify a pattern caused by vanishing density of states sites in pristine ribbons with armchair borders. 
    Impurities located on these sites become practically invisible to the interaction, but can be made accessible by a gate voltage or doping. We also demonstrate that the coupling between impurities can be turned on or off using this characteristic, which may be used to control the transfer of information in transistor-like devices.

\end{abstract}

  \maketitle 
  
    One of the continuing challenges of the present day is to increase the ratio of computational speed to power consumption and cost. 
    The field of spintronics attempts to tackle this problem by utilizing electron spin in solid state devices. 
    In contrast to charge currents, spin currents are in principle transported with substantially less dissipation\cite{taniguchi_dissipation_2014}, thus giving rise to a potentially efficient way of transporting information through nanoscale systems. 
    Development of these ideas has been frustrated by the difficulty in establishing long range spin currents, as they tend to decay over moderately short distances \cite{Zhang:2013ic}.
    These obstacles appear to be addressed by graphene and some of its allotropes, which possess extremely long range spin coherence lengths, owing to their weak spin-orbit and hyperfine interactions\cite{huertas-hernando_spin-orbit_2006,yazyev_hyperfine_2008,Dlubak:2012ib,Phillips:2016bo}. 
    
    Because pristine graphene is non-magnetic, most studies of the magnetic response of graphene are done in samples with added magnetic dopants\cite{yazyev_emergence_2010,vozmediano_local_2005,dugaev_exchange_2006,saremi_rkky_2007, brey_diluted_2007,sherafati_rkky_2011,black-schaffer_rkky_2010,uchoa_kondo_2011,black-schaffer_importance_2010,power_electronic_2011,sherafati_analytical_2011,power_strain-induced_2012,power_dynamic_2012,
      venezuela_emergence_2009-1,peng_strain_2012,gorman_rkky_2013,power_indirect_2013,duffy_variable_2014}. The overall behavior of magnetically-doped graphene is thus crucially dependent on how these dopants interact with each other. 
    Graphene plays a key role in determining the nature and range of this interaction. This is because, at all but the closest ranges, the interaction between magnetic impurities is mediated by conduction electrons of the host material. When calculated using second order perturbation theory, this is called the RKKY interaction, although the moniker is often applied to the interaction as a whole. The interaction is generally related to the dimensionality of the system, but the unusual band structure of graphene means that it sports a shorter than normal range\cite{power_electronic_2011,sherafati_analytical_2011,black-schaffer_rkky_2010,brey_diluted_2007,saremi_rkky_2007}, and as such, methods of lengthening the interaction are sought. 

    Among the possible strategies to extend the range of the magnetic interaction between impurities, 
      one that seems particularly viable is setting the impurities in precessional motion\cite{heinrich_dynamic_2003}. 
    In fact, the RKKY interaction between localized magnetic moments embedded in graphene has been predicted to become more long ranged once the magnetic moments are taken out of equilibrium and set to precess through spin pumping\cite{power_dynamic_2012}. 
    This can be understood in terms of the spin currents emanating from precessing moments. In this case, angular momentum from the moving magnetization is transferred to the conduction electrons, 
    creating a spin disturbance that propagates throughout the conducting material\cite{tserkovnyak_spin_pumping_2002,heinrich_dynamic_2003,Guimaraes:2011ks}.
    In the case of graphene, the particularly weak spin-orbit coupling of carbon drives this disturbance further afield, explaining why the range of the dynamic form of the RKKY interaction is so much more long ranged.

    Edged graphene has been highly studied for many of the same reasons as its bulk counterpart, along with some interesting features of its own \cite{acik_nature_2011,yazyev_emergence_2010,girit_graphene_2009,son_energy_2006, han_energy_2007,power_model_2009,klinovaja_rkky_2013,duffy_variable_2014}. 
    It is of particular interest to us, as its compactness makes it more likely to be of use in the channels and interconnects of future spintronic devices. 
    Despite this, the study of magnetism in edged graphene has been a little reserved, attracting little attention outside of the anti-aligned magnetic edges of zigzag nanoribbons. 
    This has likely been due to the difficulty in manufacturing well defined edges\cite{acik_nature_2011}, and to the disappearance of many interesting properties of nanoribbons in the presence of disorder\cite{kunstmann_stability_2011}. 
    However, there has been increasing success in producing well defined edges in recent years\cite{zhang_experimentally_2012,wang_etching_2010,kosynkin_longitudinal_2009}, which will likely reinvigorate some research in this area. 
    Indeed, the RKKY coupling between magnetic moments in nanoribbons has been recently investigated\cite{duffy_variable_2014}. 
    Interestingly, it has been shown that the static form of this interaction actually alternates between long- and short-ranged depending on the distance of the magnetic dopants to the graphene edge. 
    This naturally raises the question of whether a similar pattern is to be seen in the dynamic version of this interaction, wherein the magnetic moments interact by means of collective excitations. 
    The dynamic interaction possesses some obvious benefits above its static counterpart, being, in most cases, both stronger and longer ranged\cite{heinrich_dynamic_2003,power_dynamic_2012}. 
    Aside from this, it is the interaction type that specifically exploits one of the key benefits of graphene, namely its long spin-coherence length. 
    Previous studies of the dynamic interaction in bulk graphene show that it is successful at mitigating its rather weak static coupling \cite{power_dynamic_2012} and that this  interaction can be tuned using potential ``lenses'' \cite{guimaraes_graphene_2011}.
	
    Given the current state of the field, we perform a study on the nature of the dynamic interaction in edged graphene. 
    Although we focus our examination in armchair-edged nanoribbons with top adsorbed impurities, most of our results are present in any 1D metallic host connected to magnetic units. 
 We also constrict the investigation for ferromagnetic interactions of impurities adsorbed to carbon atoms belonging to the same sublattice to avoid competition between external magnetic field and exchange interaction. It is worth noticing that certain impurities prefer to occupy one of the two sublattices rather than being randomly dispersed in graphene  \cite{lawlor_friedel_2013,lawlor_sublattice_2014,Lv:2012ji,ZabetKhosousi:2014ct,Lin:2015gq,Usachov:2016eb}. 
 
 The focus of this paper will mainly be on the lifetime and character of the spin excitations, 
    quantities that are commonly measured by ferromagnetic resonances\cite{heinrich_dynamic_2003} but may also be probed using transport methods \cite{Dlubak:2012ib,costache_electrical_2008,tombros_electronic_2007}. By establishing the characteristics of these quantities and how they behave in edged graphene, we hope to provide a way in which the magnetic excitation response of magnetically-doped nanoribbons can be engineered. 
 
  \section{Theory and Model}
  
\begin{figure}
      \centering
      \includegraphics[width=0.45\textwidth]{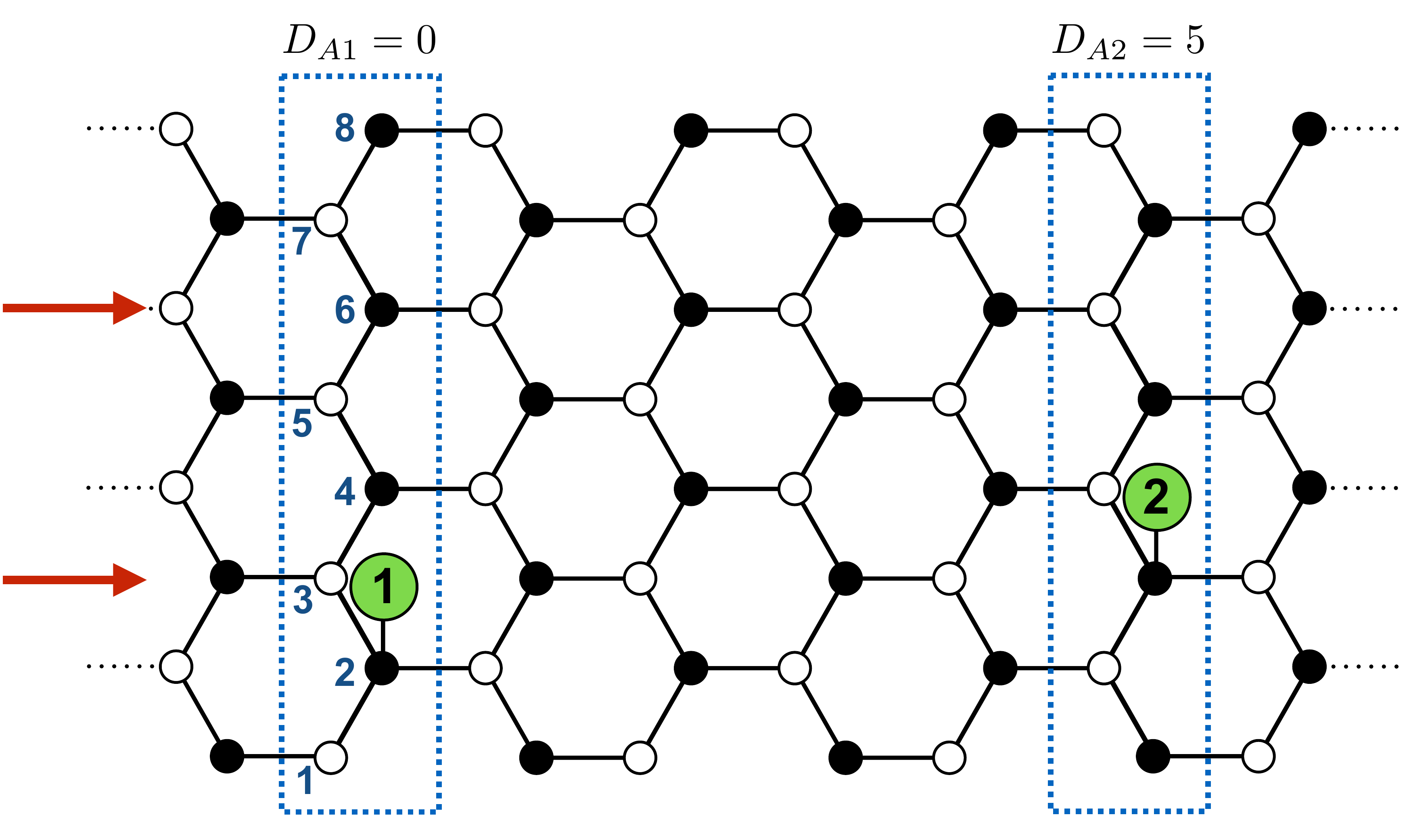}
      \caption{A schematic diagram of a graphene nanoribbon with $N$ atoms across its width. The lattice parameter is given by $a$.
	Impurities 1 and 2 are shown as numbered circles. The location of impurity $n$ is identified by the labels $D_{An}$ and $D_{Zn}$, which denotes the unit cell depicted by the (blue) dashed rectangles and the carbon site inside this cell, respectively. The separation along the edge direction is given $D_A = D_{A1} - D_{A2}$.
	Here, $D_{Z1}=2$, $D_{Z2}=3$, $D_A=5$ and $N=8$. The (red) arrows indicate lines with vanishing local density of states.}
      \label{System}
    \end{figure}

    To illustrate our notation we make use of the sample system with two top-adsorbed impurities shown in Fig. \ref{System}. The ribbon size is determined by the number of atomic sites in a zigzag ``strip'', given by $N$.
    Ribbons of width $N=(3n+2)$ where $n \in \mathbb{N}$ are metallic, all other ribbons are semiconductors, although a metallic character may be achieved by a gate voltage or doping.
    Since we wish to examine the dynamic interaction in the presence of spin waves, we will work exclusively with metallic ribbons.
    The vertical positions of the impurities are denoted with respect to a line below the sheet located where the next atomic site would be. 
    The vertical positions are given by $D_{Z1}$ and $D_{Z2}$, and the horizontal separation by $D_A$. 
    The units denoting vertical position are given in $\frac{a}{2}$, so that they essentially count atomic positions from the edge.
    Similarly, the units of horizontal separation are given in $\sqrt{3} \frac{a}{2}$ and so count atomic sites in the horizontal direction.
     In the model system shown in Fig.~\ref{System}, our two top-adsorbed impurities are labeled 1 and 2. 

It is worth mentioning that in pristine armchair ribbons the local density of states (LDOS) at the Fermi level ($E_F$) vanishes for atomic sites labeled by $D_Z = 3n$. We refer to the periodic lines across the nanoribbon comprising these sites as vanishing LDOS (VLDOS) lines, and we shall see that impurities adsorbed to such sites acquire rather unique characteristics. Although VLDOS is normally seen in armchair-edged ribbons, LDOS modulation across the width of any metallic ribbon is a common feature.

    The electronic structure of our system is described by the model Hamiltonian $\hat{H} = \hat{H}_0 + \hat{H}_{\text{imp}}$, where $\hat{H}_0$ characterizes the ``pristine'' lattice, and $\hat{H}_{\text{imp}}$ accounts for the adsorbed magnetic impurities. Here, 
    \begin {equation}\label {H0}
\hat{H}_0 = \sum_{i,j,\sigma} t_{ij} \hat{c}^{\dagger}_{i \sigma} \hat{c}_{j \sigma}\,, 
\end{equation}
and
\begin{equation}\label{himp}      
    \hat{H}_{\text{imp}} = \sum_{l,\sigma} \epsilon_l  \hat{n}_{l \sigma} + \sum_{i,l,\sigma} t'_{il} \hat{c}^{\dagger}_{i \sigma} \hat{c}_{l \sigma} + \hat{H}_{\text{int}} \,,
    \end{equation}
where
\begin{equation}\label{hint} 
   \hat{H}_{\text{int}} = 
    U \sum_{l,\sigma}  \hat{n}_{l \sigma} \hat{n}_{l \bar{\sigma}} + g \mu_B B_0 \sum_l \hat{S}^z_l. 
    \end{equation}
    $\hat{c}^{\dagger}_{i \sigma}$ and $\hat{c}_{j \sigma}$ designate the creation and annihilation operators for electrons with spin $\sigma$ at sites $i$ and $j$, respectively. $t_{ij}=t$ is the nearest-neighbor hopping integral, which we take as the energy unit throughout the paper. 
    The pristine system is simulated in the tight-binding representation, where we only consider the interaction between $\pi_z$ orbitals on nearest neighbors, which is known to provide a good approximation to the band structure of graphene. $\hat{H}_{\text{imp}}$ describes the impurities adsorbed to the carbon nanoribbon; $l$ labels the impurity site, $\epsilon_l$ represents the impurity atomic energy, and $\hat{n}_{l \sigma}= \hat{c}^{\dagger}_{l \sigma}\hat{c}_{l \sigma}$ is the corresponding local electronic occupation number operator. $\hat{H}_{\text{int}}$ describes the effective Coulomb interaction between two electrons on the same impurity site $l$, plus an on-site static Zeeman interaction between an external magnetic field $B_0$ and the magnetization of the impurity which serves to create a default magnetic alignment arrangement. For simplicity we take $\epsilon_l = 0$, the nearest-neighbor hopping integrals between the impurities and the carbon sites $t'_{i,l} = t$, and the Zeeman energy $g\mu_B B_0 = 10^{-3}t$.
    
    First we calculate analytically the one-electron Green functions (GF) for the nanoribbon characterized by  $\hat{H}_0$\cite{power_electronic_2011}, and then add the magnetic impurities described by  $\hat{H}_{\text{int}}$ using Dyson's equation. The magnetic ground state is determined self-consistently \cite{barbosa_spin_2001}.
    Finally, we consider a time-dependent oscillatory transverse magnetic field $\mathbf{h}_\bot$ given by 
    \begin{equation} \mathbf{h}_\bot = h_0 [ \cos(\omega t) \hat{x} - \sin(\omega t) \hat{y}]\,, \end{equation} 
    that sets the impurities local magnetic moments into precession. 
    The effects of this time-dependent external perturbation Hamiltonian are calculated within linear response theory. 
    The dynamic transverse spin susceptibility is given by 
    \begin{equation} 
      \chi_{m,l}^{+ -}(t) = -\frac{i}{\hbar} \Theta (t) \expec{\left[\hat{S}^+_m(t),\hat{S}^-_l(0) \right]} 
    \end{equation}
    where $\hat{S}^+$ and $\hat{S}^-$ are the spin raising and lowering operators, and the non-local transverse spin susceptibility  $\chi_{m,l}^{+ -}(\omega)$ represents the response of the system at site $m$ due to the time dependent magnetic field field applied at site $l$. 
    This quantity contains the full description of the change in the spin density of the system.
    Its Fourier transform $\chi_{m,l}^{+ -} (\omega)$ gives the response of the system to harmonic excitations of a given frequency $\omega$. 
    In our approach, this quantity is calculated within the random phase approximation (RPA), which is limited to low temperature but sufficient to describe the essence of the spin excitations of our system.  
    
    The spectral density, related to the imaginary part of the local response function $S_m(\omega) \equiv \operatorname{Im}\chi^{+-}_{m,m}(\omega)$, describes the spin excitations of the system.
    In principle one would look for poles in $S_m(\omega)$ in order to find the spin excitation energies. However, for magnetic
    objects in contact with nonmagnetic metallic substrates, $S_m(\omega)$ does not have true poles but only
    broadened resonances. This stems from the fact that spin excitations are not exact eigenstates of those systems, but may decay
    into Stoner excitations~\cite{barbosa_spin_2001}. This interpretation is strongly suggested by the form of the RPA equation that relates $\chi^{+-}(\omega)$ to its mean-field version $\chi^{0+-}(\omega)$, schematically given by 
    \begin{equation}
     \label{RPAschematic}
      \chi^{+-}(\omega) = \frac{\chi^{0+-}(\omega)}{1+\chi^{0+-}(\omega)U} ,
    \end{equation}
    where an arbitrary matrix element of the mean-field susceptibility is given by
\begin{equation}\label{chi0}
\begin{split}
\chi^{0+-}_{m,l}(\omega) = -\frac{\hbar}{4\pi}\int^{E_F} dE &\left\{G^{\downarrow}_{ml}(E+\omega+i0^+) \Im G^{\uparrow}_{lm}(E)\right.\\
&\left.+\Im G^{\downarrow}_{ml}(E) G^{\uparrow}_{lm}(E-\omega-i0^+) \right\}\ .
\end{split}
\end{equation}
Here $G^{\sigma}_{ml}(E)$ are the monoeletronic propagators for electrons with spin $\sigma$ between sites $m$ and $l$, and we have defined $\Im G^{\sigma}_{ml}(E)=\frac{1}{2i}\left[G^{\sigma}_{ml}(E+i0^+)-G^{\sigma}_{ml}(E-i0^+)\right]$.
    The transverse spin susceptibility displays resonances with linewidths determined by $\mathrm{Im}[\chi^{0+-}(\omega)]$,
    which is the spectral density of Stoner (single-particle spin-flip) excitations.
    The projection of these modes onto the adatom is correlated to the density of Stoner modes at the attachment site. This can be seen analytically by inspecting Eq. \ref{chi0}. The local Green function on the adatom carry the full information about its connection with the substrate, thus tying inextricably the densities of Stoner modes in the adatom and in the substrate.
    It has been shown that the imaginary part of the non-interacting susceptibility increases linearly with $\omega$ in the limit $\omega \rightarrow 0$, and its rate of change is given by the product of the electronic densities of states for majority and minority spins at the Fermi energy $E_F$, i.e., $\lim_{\omega\rightarrow0}\operatorname{Im}[\chi^{0+ -}_{mm}(\omega)] \sim \omega\,\rho_\uparrow(E_F)\rho_\downarrow(E_F)$ \cite{PhysRev.160.590,lounis_dynamical_chi_prb_2015}. In most metallic systems $\mathrm{Im}[\chi^{0+-}(\omega)]\ne 0$,
    and collective spin excitations usually have finite lifetimes. There are very particular situations, however, for which 
    some collective excitations may ``decouple'' from the Stoner continuum, as we shall discuss later. 
    
    In this paper we consider the magnetic objects to be transition metal atoms adsorbed to graphene nanoribbons. Thus, it is suitable to work in real space, and use the atomic site representation of the transverse susceptibility presented in Eq.~\ref{RPAschematic} above. Within the single 
    band Hubbard model given by Eqs.~\ref{H0}-\ref{hint}, the RPA equation for the susceptibility reads
    \begin{equation} 
      \chi_{m,l}^{+ -}(\omega) = \chi_{m,l}^{0+ -}(\omega) - \sum_{l'}\chi_{m,l'}^{0+ -}(\omega)U_{l'}\chi_{l',l}^{+ -}(\omega),
      \label{RPAmatrixEq}
    \end{equation}
    where we assume $U_l\ne 0$ at the adatoms sites only. Clearly, in the single impurity case, the solution of the RPA equation for $ \chi_{1,1}^{+-}$, where $1$ labels the adatom site, takes the form of Eq.~\ref{RPAschematic}. 
    When more than one magnetic adatom is present, Eq.~\ref{RPAmatrixEq} can be cast into matrix form and solved accordingly, $[\chi^{+-}] = (1+[\chi^{0+-}][U])^{-1}[\chi^{0+-}]$, where the quantities in square brackets represent matrices in real space including all the magnetic adatoms.
       
  \section{Single Impurities}
  
\begin{figure}
      \centering
      \includegraphics[width=0.5\textwidth]{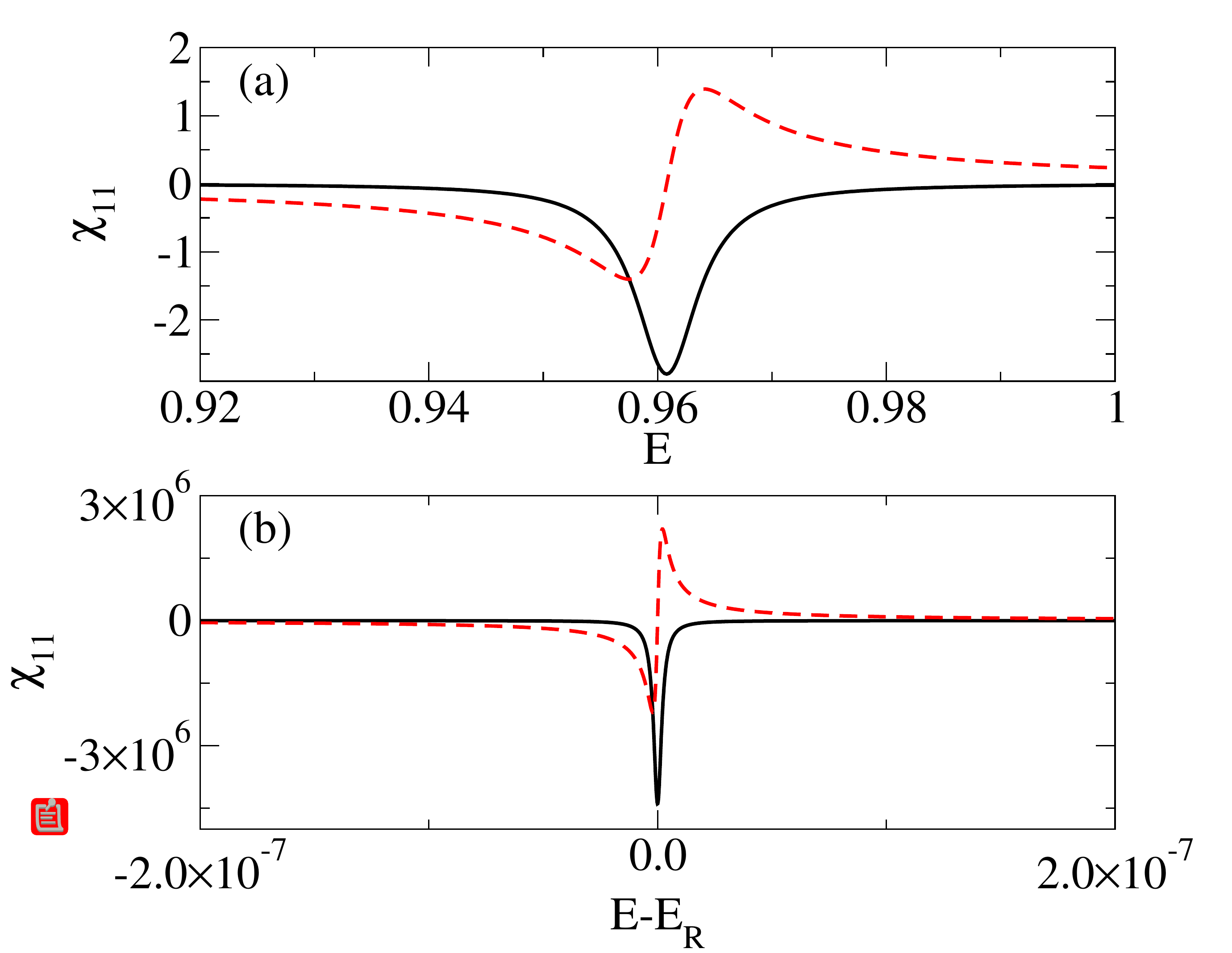}
      \caption{Real and imaginary parts of $\chi^{+-}_{1,1}$ (in arbitrary units) as functions of the energy $E=\hbar \omega$ (in units of $10^{-3}t$), where $\omega$ represents the frequency of the oscillatory field. For clarity, the bottom panel is plotted as a function of $E-E_R$, where $E_R = 0.9639127436t$. The results are for a single impurity: (a) adsorbed to the edge of a nanoribbon ($N=5$, $D_Z=1$); (b) adosorbed to a VLDOS site ($N=5$, $D_Z=3$). Re$\chi^{+-}_{1,1}$ (red dashed lines), and Im$\chi^{+-}_{1,1}$ (black solid lines).} 
      \label{X1}
    \end{figure}

    We begin by examining the spin susceptibility for single impurities. The onsite spectral density $S_{1}(E)$, where $E=\hbar \omega$, and $\omega$ is the angular frequency of the driving field, gives clues as to the lifetime and character of the excitation.
    Fig. \ref{X1}(a) shows the real and imaginary parts of $\chi^{+-}_{1,1}$ as functions of the energy $E$ for a single impurity at the edge of a nanoribbon.
    The resonance peak describes the precession mode of the local magnetic moment activated by the time-dependent external perturbation, and it  occurs, as expected, around the Larmor frequency $g \mu_B B_0$. It is noteworthy that by calculating $\chi^{+-}_{1,1}$ we are focusing on the local spin dynamics, where the magnetic moment of the impurity resides. In this case, as discussed in Refs.~\onlinecite{PhysRev.160.590, PhysRevB.68.224414}, a spatially localized probe will detect, in principle, a relatively broad structure that may be substantially shifted from the free-electron resonance frequency, corresponding to g=2. Both the broadening and the shift depend on how the impurity atom will couple to the substrate's particle-hole pairs. In  Fig. \ref{X1}(a) it is clear that the real part of $\chi^{+-}_{1,1}$ crosses zero while the imaginary part has a maximum, confirming that this is an electronic collective mode.
The linewidth of the peak is non-zero, which means that the lifetime of the excitation (given by the inverse of the linewidth) is finite. The origin of this relaxation is the decay of the collective mode into single-particle Stoner excitations \cite{PhysRev.160.590,PhysRevB.68.224414}. As a result of this process, spin angular momentum is transferred from the magnetic adatom to the non-magnetic substrate. This mechanism, known as \textsl{spin pumping}, may be used to inject a pure spin on adjacent materials\cite{tserkovnyak_spin_pumping_2002}. Further calculations show that broadened peaks are ubiquitous at sites with non-VLDOS.
  
    Fig. \ref{X1}b shows the same calculation for an impurity adsorbed to a site belonging to the line of VLDOS labeled by $D_Z=3$.
   Although the real and imaginary parts are similar in shape to the previous case, the resonance peak is orders of magnitude larger and narrower than the previous case, resembling a delta function. For that reason, the figure was plotted relative to the resonance energy $E_R$. This indicates that the excitation has an extremely long lifetime.
    The mechanism underlying such behavior is the VLDOS at the Fermi level in the host site atop which the impurity is placed. In this case, the coupling between the magnetic adatom and the particle-hole pairs in the substrate is strongly reduced. Therefore, since there are no possible Stoner modes for the excitation to decay into, the magnetization of the adatom behaves essentially as an isolated spin. Thus, because our model has rotational symmetry in spin space, the precession of a virtually isolated magnetic moment is not damped, hence the linewidth associated with this excitation practically vanishes.
\begin{figure}
\subfloat{%
  \includegraphics[width=0.45\textwidth]{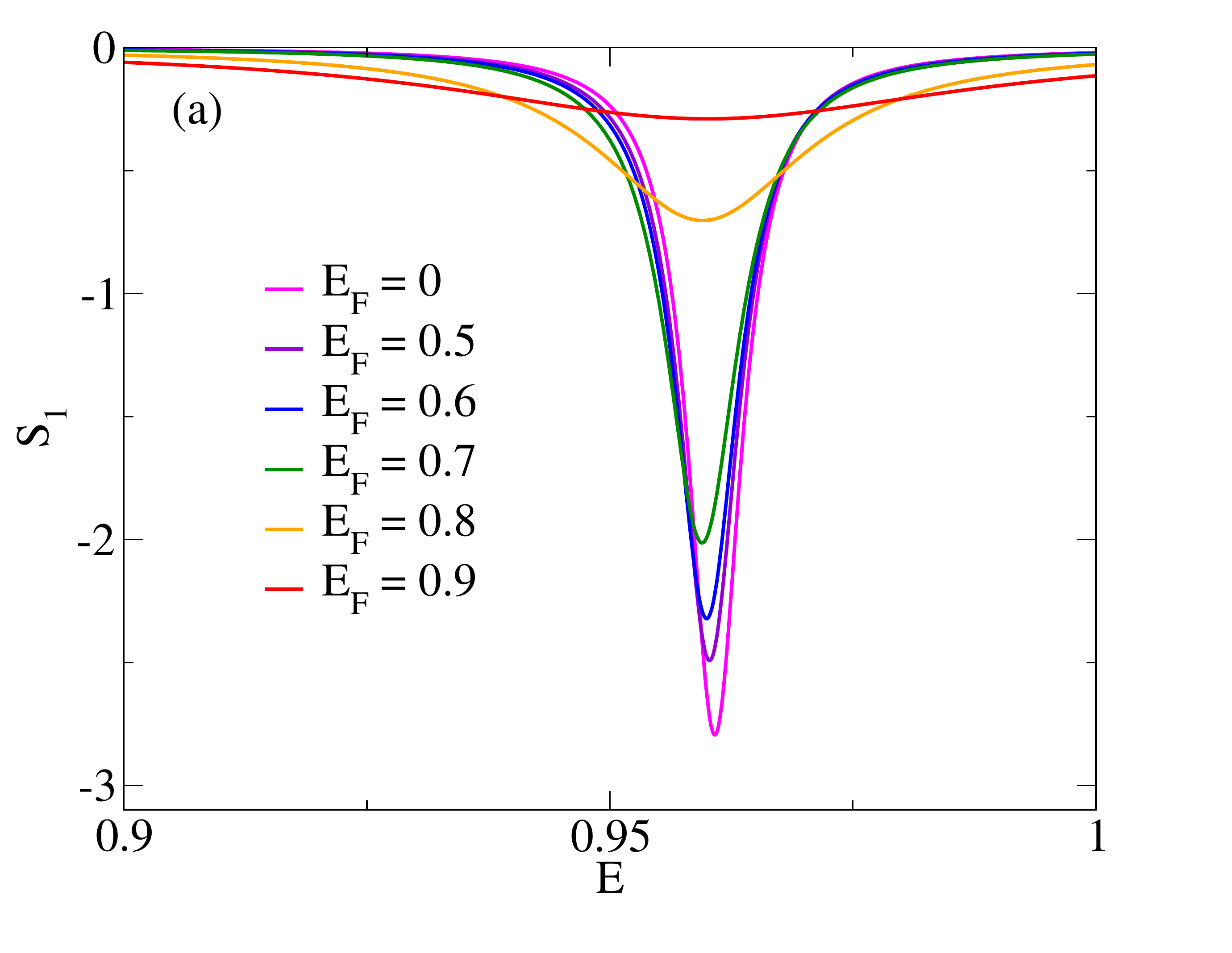}%
}\hfill\\
\subfloat{%
  \includegraphics[width=0.45\textwidth]{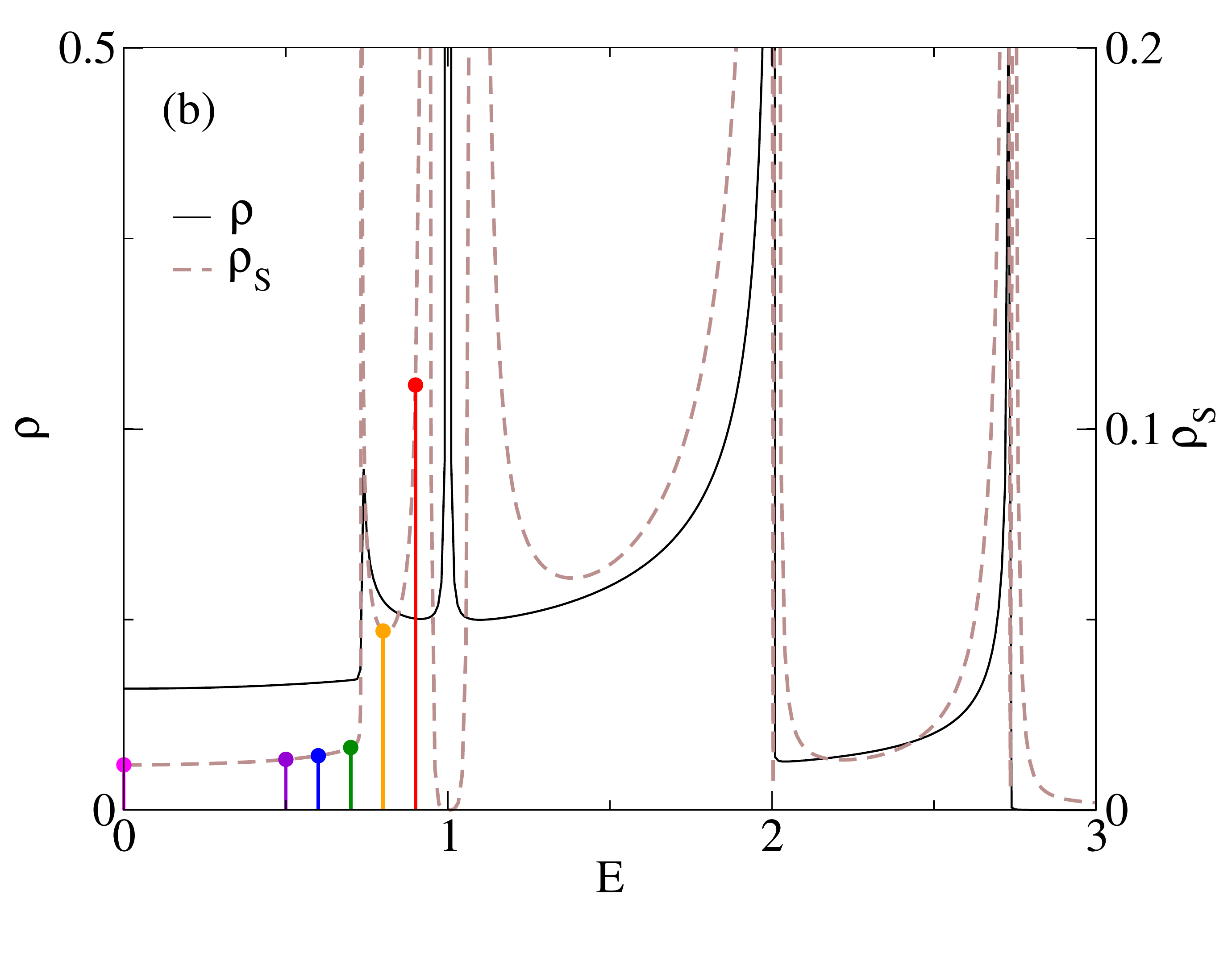}%
}
\caption{(a) Spectral density $S_{1}$ calculated as a function of energy $E=\hbar \omega$ (in units of $10^{-3}t$) for a single impurity, and different values of the Fermi energy $E_F$. Each color corresponds to a specific value of $E_F$. The impurity is adsorbed to a site in the edge of a nanoribbon with $N=5$, along the line labeled by $D_Z=1$.
 (b) Local density of states ($\rho$) at the pristine-ribbon site where the impurity is adsorbed to (black curve) as a function of the energy, and local density of Stoner modes ($\rho_S$) at the magnetic impurity site (brown curve) as a function of $E_F$. Color marks indicate the values of $E_F$ where the spectral densities depicted in panel (a) were calculated. The LDOSM $\rho_S(E_F)$ is represented by $\lim_{\omega\rightarrow0}\mathrm{Im} \chi_{11}^0 (\omega)/\omega$.}
\label{XvEF}
\end{figure}

We now show that it is possible to tune the lifetime of the excitation by changing the Fermi level of the system, something which may be experimentally achieved with the use of a gate voltage or by doping. First we shall examine a non-VLDOS case where a single impurity is located at the edge of a nanoribbon with $N=5$, along the line corresponding to $D_Z=1$. The results are depicted in Fig. \ref{XvEF}(a) which displays the spectral density $S_{1}(E)$ plotted against the energy $E=\hbar\omega$, for different values of $E_F$. We perceive that as $E_F$ increases initially there is little change in the peak shapes, but beyond a threshold the peak becomes smaller and broader, indicating a strong reduction in the lifetime of the excitation. One is naturally led to correlate this behavior with the LDOS of the host site at $E_F$ ($\rho(E_F)$) atop which the impurity is adsorbed. In Fig. \ref{XvEF}(b) we see that $\rho(E_F)$ exhibits a plateau whose border is found at a value of $E_F$ slightly greater than 0.6, followed by a sharp peak. We may argue that the greater the LDOS, the larger would be the density of Stoner modes for the excitation to decay into, leading to a stronger damping. In fact, under certain conditions it is possible to show that the linewidth of the local spin excitation is indeed proportional to $\left(\rho(E_F)\right)^2$, while the $g$ shift varies linearly with $\rho(E_F)$ \cite{PhysRev.160.590}. 
However, although this is verified in several situations, it is not always true. What primarily determines the damping of the spin excitation mode is the local density of Stoner modes (LDOSM) at the impurity site, not the local density of electronic states where the impurity will be attached to. To illustrate this point, in    
Fig. \ref{XvEF}(b) we compare  $\rho(E)$ with the LDOSM ($\rho_S (E_F)$) at the impurity site, both calculated as functions of energy. Since the general behavior of the LDOSM is linear at this energy scale, $\rho_S (E_F)$ is represented by its slope at zero frequency, $\lim_{\omega\rightarrow0}\mathrm{Im} \chi_{11}^0 (\omega)/\omega$.  
The color marks indicate the corresponding values of $E_F$ for which the susceptibilities displayed in Fig.~\ref{XvEF}(a) were calculated.
A close inspection of those figures  shows that the linewidths ($\Gamma$) of the spectral functions depicted in Fig.~\ref{XvEF}(a) follow the behavior of the LDOSM. We note that $\Gamma$, measured by the full width at half maximum of the spectral function, is inversely proportional to $\rho_S(E_F)$, but not always to $\rho(E_F)$. For instance, when $E_F$ changes from 0.8 to 0.9, the value of $\rho(E_F)$ slightly decreases, whereas $\rho_S(E_F)$ increases substantially, and so does $\Gamma$. Even more remarkable is the case when $E_F \approx 1$ which corresponds to a van Hove singularity. In this situation, we see that $\rho(E)$ is divergent, whereas $\rho_S(E)$ goes to zero, as illustrated in Fig.~\ref{XvEF}(b).
In fact, we have found that the linewidth of the spin excitation actually goes to zero for $E_F = 1$, in agreement with the behavior of $\rho_S$. The corresponding narrow spectral line is not shown in panel (a) because the $g$-shift for $E_F = 1$ is substantially larger.

\begin{figure}
      \centering
      \includegraphics[width=0.5\textwidth]{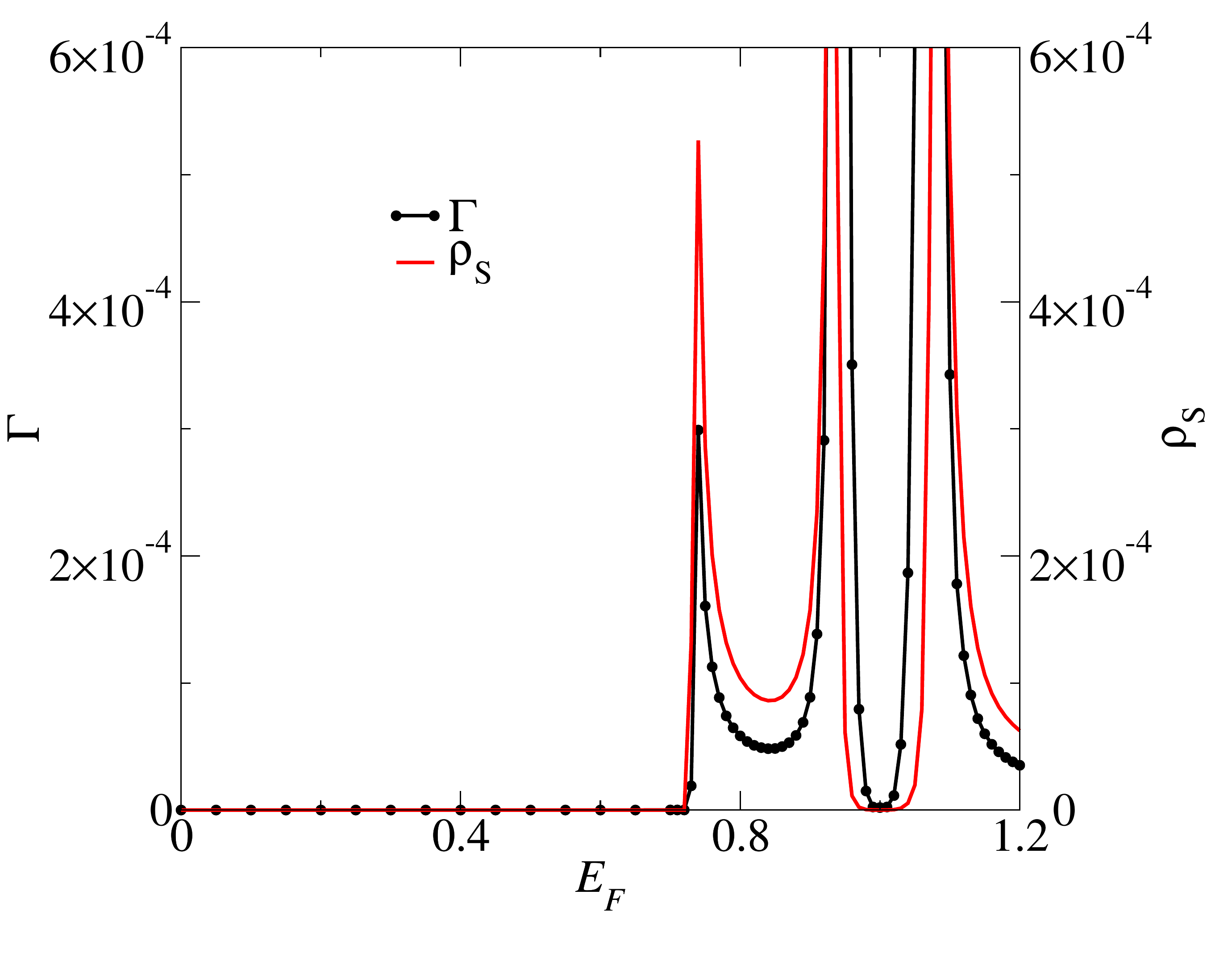}
      \caption{Full width at half maximum $\Gamma$ (black curve) and local density of Stoner modes $\rho_S$ (red curve) calculated as a function of the Fermi energy $E_F$ for a single impurity at the VLDOS site $D_Z=3$ in a nanoribbon of width $N=5$. Energies are expressed in units of the nearest-neighbor hopping integral $t$. The LDOSM $\rho_S$ is represented by $\lim_{\omega\rightarrow0}\mathrm{Im} \chi_{11}^0 (\omega)/\omega$.} 
      \label{FWHMvEF_DZ3}
    \end{figure}

Similar results are obtained for the case in which the impurity is placed atop a VLDOS. In Fig. \ref{FWHMvEF_DZ3} we show the spectral linewidths $\Gamma$ calculated as a function of $E_F$ for a single impurity adsorbed to a site situated along $D_Z=3$ in a nanoribbon of width $N=5$. As $E_F$ exceeds the gap region, $\Gamma$ raises sharply, and begins to vary, taking relatively high and low values as $E_F$ increases, closely following the behavior of $\rho_S(E_F)$, which is also portrayed in the same figure. Once again we see that $\Gamma$ vanishes at the van Hove singularity at $E_F=1$, as so does $\rho_S$. 
  We note that our results have been obtained for a very narrow ribbon that displays a relatively large central plateau, whose edges may be difficult to reach by either gating or chemical doping. Nevertheless, they certainly also apply to wider strips, with much narrower plateaus, since the plateau broadness decreases with the ribbon width.  

        \begin{figure}
      \centering
      \includegraphics[width=0.5\textwidth]{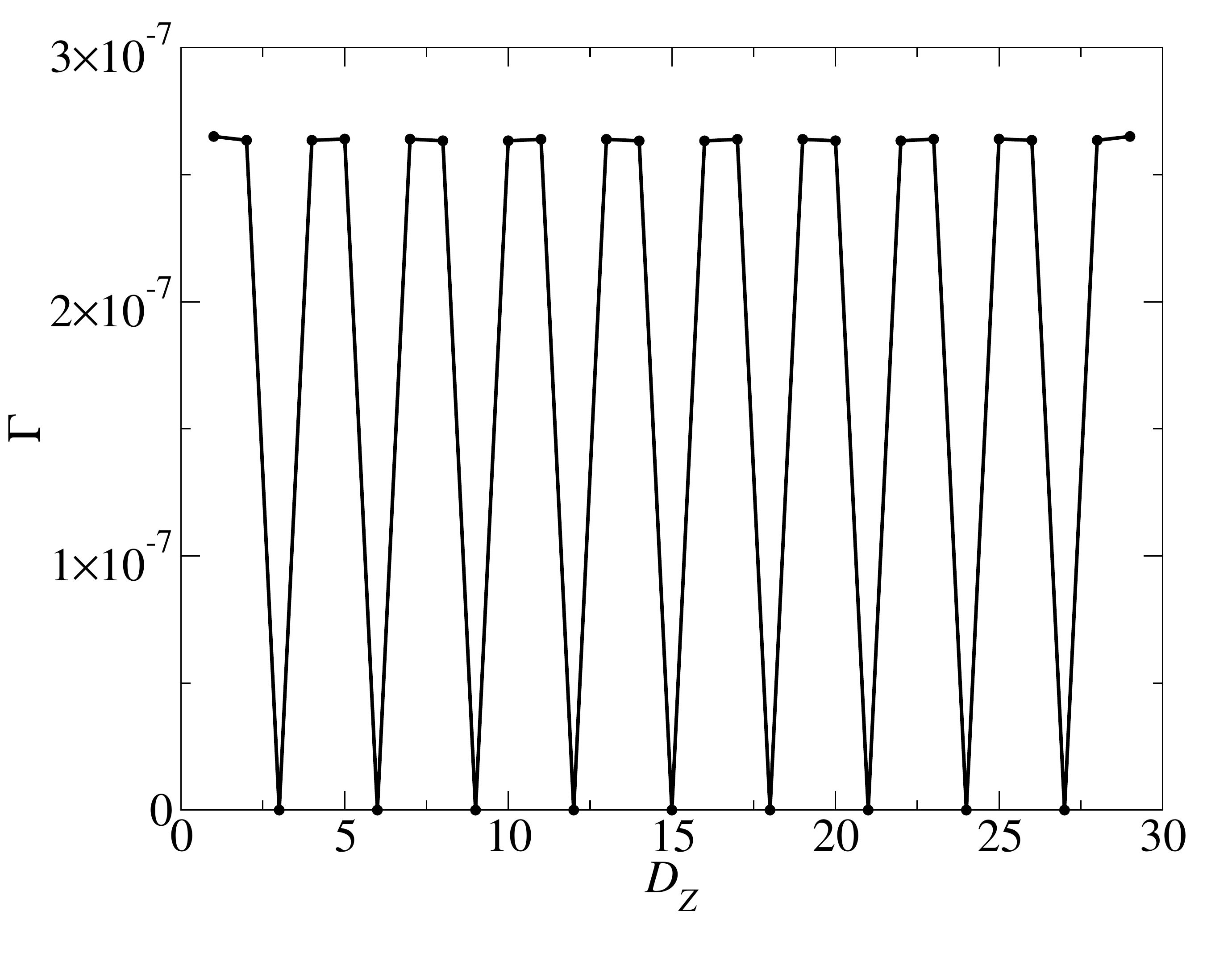}
      \caption{Full width at half maximum $\Gamma$ of the spectral-lines peaks for impurities adsorbed at different distances ($D_Z$) from edge of a nanoribbon with width $N=29$. 
	The FWHM reduces almost to zero for impurities adsorbed onto sites belonging to the VLDOS lines, indicating an extremely long-lived excitation in these cases. Energies are expressed in units of the nearest-neighbor hopping integral $t$.}
      \label{X1FWHM}
    \end{figure}

    Fig. \ref{X1FWHM} exhibits how the linewidth of the spectral function varies with the adatom's position across the nanoribbon width. It highlights its vanishing widths when the magnetic impurities are placed atop sites with VLDOS at $E_F$.
    This triadic pattern persists across the body of the nanoribbon, and has been verified for nanoribbons of varying widths. It stresses that the sensitivity of the excitations to the adsorbate position is remarkable, and it is natural to explore how this influences the interaction between multiple impurities. 
  
  \section{Two Impurities}
  
    For two magnetic impurities adsorbed to the nanoribbon there will be two kinds of indirect magnetic interactions mediated by the conduction electrons in the substrate. 
    One is the \textit{static} RKKY-like interaction, which emerges from the induced spin polarization on the substrate due to the adatoms. 
    The other is the \textit{dynamic} interaction that appears when the magnetization of one adatom is set into precession, pumping a spin current with transverse polarization into the substrate, which will be partially absorbed by the magnetic moment of the other adatom, and consequently disturbing its spin moment. 
    
\begin{figure}
      \centering
      \includegraphics[width=0.5\textwidth]{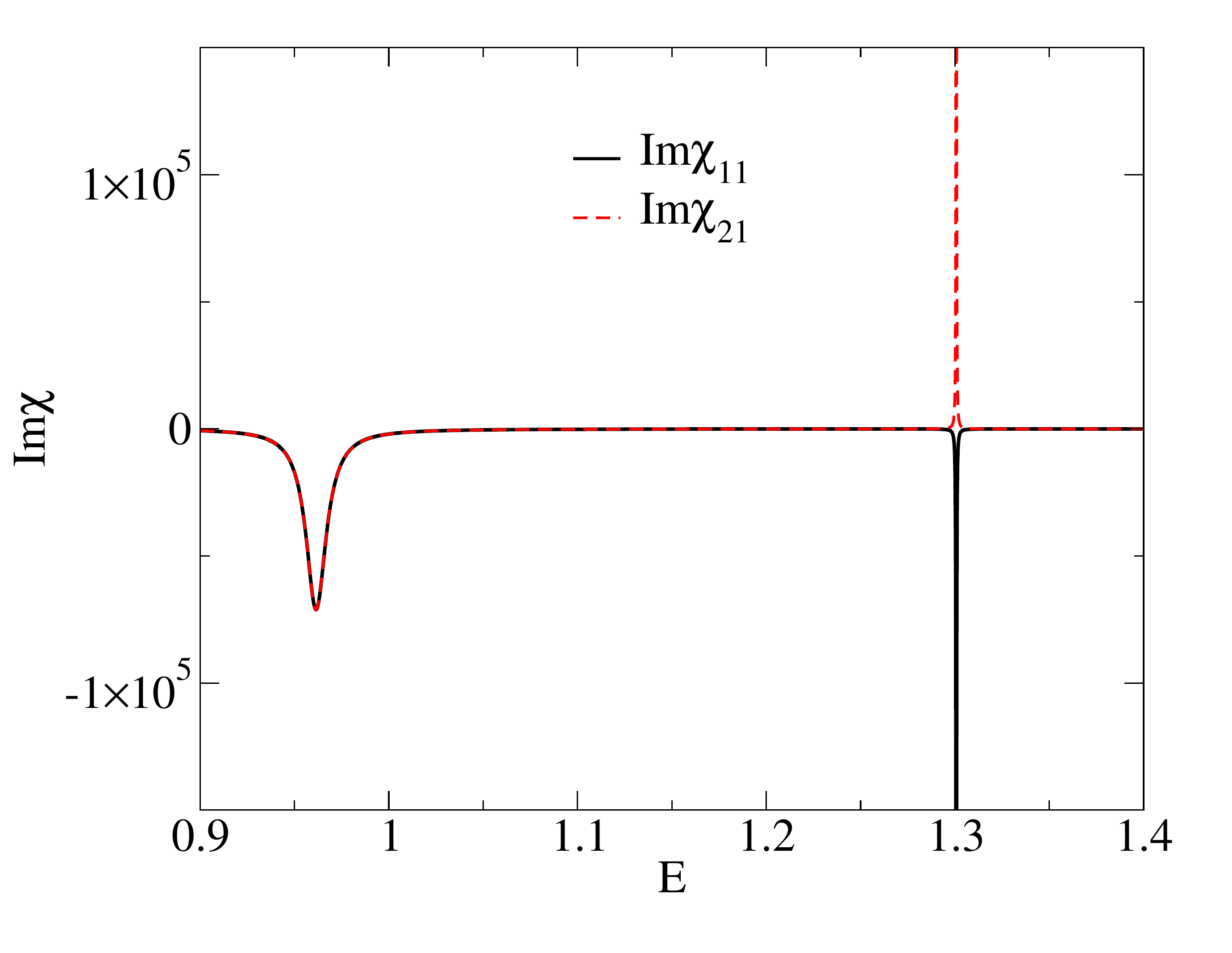}
      \caption{Imaginary parts of the local $\chi^{+-}_{11}(\omega)$ (solid black line), and non-local $\chi^{+-}_{21}(\omega)$ (dashed red line) transverse spin susceptibilities as functions of the energy $E=\hbar \omega$ (in units of $10^{-3}t$), for two impurities (1 and 2) adsorbed to non-VLDOS sites of a nanoribbon with $N=5$. Here,  $D_{Z1}=D_{Z2}=1$, and $D_A=10$.
	The acoustic and optical modes are clearly visible, indicating that the magnetic impurities are ferromagnetically coupled, as expected.}
      \label{2XV0}
    \end{figure}

    When the adsorbed magnetic units are coupled, we expect two excitation peaks consistent with the existence of two precession modes. One in which the moments precess in phase (acoustic mode) and another where they precess out of phase (optical mode).
    These can be analyzed by comparing the imaginary parts of the diagonal and off-diagonal matrix elements of the transverse spin
    susceptibility, $\operatorname{Im}\chi^{+-}_{11}(\omega)$ and $\operatorname{Im}\chi^{+-}_{21}(\omega)$. 
  The results are depicted in Fig. \ref{2XV0}, where we show both quantities calculated for a typical case in which the impurities are adsorbed to non-VLDOS sites. We recall that $\operatorname{Im}\chi^{+-}_{11}(\omega)$ and $\operatorname{Im}\chi^{+-}_{21}(\omega)$ specify the corresponding transverse spin components acquired by adatoms $1$ and $2$, respectively, due to a time-dependent oscillatory field applied to the adatom $1$. In Fig. \ref{2XV0} it is clear that the first peak has the same phase and intensity in both adatoms, while the second changes sign in  $\operatorname{Im}\chi^{+-}_{21}(\omega)$. This indicates that for the first peak the phase difference between the magnetization precession of adatom $2$ with respect to that of adatom $1$ is zero, whereas for the second peak the phase difference is $\pi$. These resonances are usually referred to as acoustical and optical precession modes, and the difference between their resonance energies is a measure of the exchange coupling linking the two magnetic impurities.
  
    \begin{figure}
      \centering
      \includegraphics[width=0.5\textwidth]{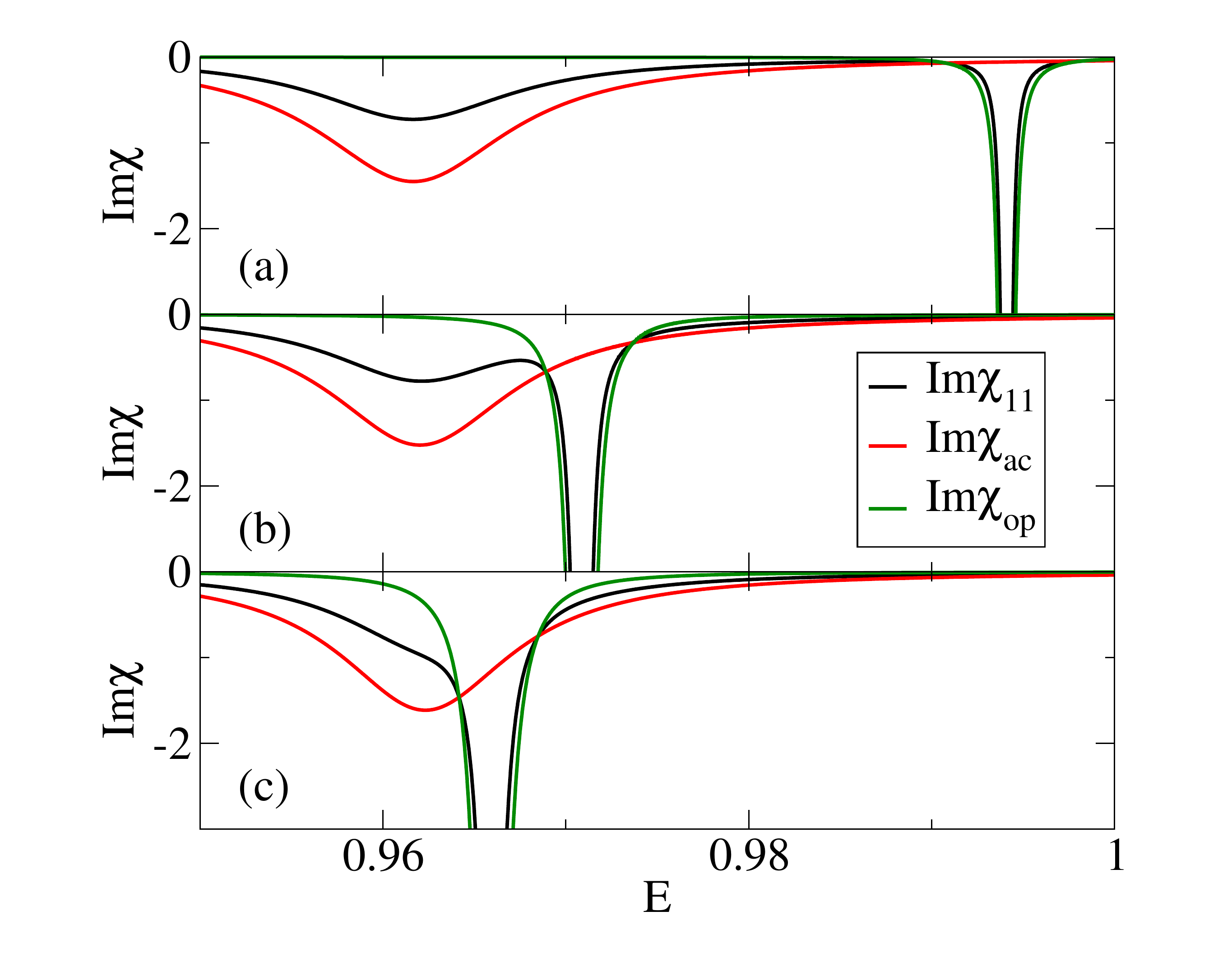}
      \caption{Imaginary part of the local transverse spin susceptibility  $\chi^{+-}_{11}(E)$ (solid black line) calculated as functions of the energy $E=\hbar \omega$ (in units of $10^{-3}t$), for $D_{Z1} = D_{Z2} = 1$. (a) $D_A = 100$, (b) $D_A = 300$, and (c) $D_A = 500$. The red and green lines represent the acoustical and optical components of $\chi_{11}$, respectively. }
      \label{snapshots}
    \end{figure}

A distinct feature in this case is the severely reduced width of the optical mode in comparison with the acoustic one. This is highly unusual, since in most metallic systems the density of Stoner modes increases with energy. However, since the adatoms are adsorbed to two equivalent sites the spin excitations of a given parity can only decay into Stoner modes with the same parity. Thus, the relevant quantity to determine the lifetime of the optical mode is the equivalent optical Stoner mode density. 

In order to inquire into the nature  of the two modes and fully identify their characters one may diagonalize the $2 \times 2$ susceptibility matrix to obtain the two eigenvalues and the corresponding  eigenvectors. Since in this case the two impurities occupy equivalent sites, the two eigenvalues are given by $\lambda_{\pm} = \operatorname{Im}\chi_{11} \pm  
\operatorname{Im}\chi_{21}$, corresponding to the symmetric and anti-symmetric precession modes, which hereafter we refer to as acoustical and optical combinations ($\operatorname{Im}\chi_{\text{ac}}$ and $\operatorname{Im}\chi_{\text{op}}$), respectively. Following this procedure for the non-interacting susceptibility matrix we find that the optical Stoner mode has negligible spectral density (not shown), thus leading to a very narrow linewidth of the optical spin-excitation mode. 

The same method is used to analyze the spectral functions of a pair of magnetic impurities adsorbed to non-VLDOS sites, separated by different distances $D_A$, located at the edge of a carbon nanoribbon with $N=5$ in breadth. In Fig.~\ref{snapshots} we choose three snapshots of those spectral functions for $D_A = 100, 300, 500$, that are depicted in panels (a), (b) and (c) by solid black lines, respectively. As the impurities are placed further apart, the splitting between the two modes decreases, to an extent that they start to merge, and culminate in two independent modes associated with non-interacting impurities eventually. By diagonalizing the interacting spin-susceptibility matrix we are able to separate the two modes and obtain their linewitdhs, even when the magnetic coupling between the magnetic impurities is sufficiently small for the peaks to join together. The separation into optical and acoustical is exhibited in each panel of Fig.~\ref{snapshots} by the red and green lines, respectively. They include contributions from both sites, which means that the spectral function  $\operatorname{Im}\chi_{11}$, represented by the black line, splits into $\left(\operatorname{Im}\chi_{\text{ac}} + \operatorname{Im}\chi_{\text{op}}\right)/2$. 

\begin{figure}
      \centering
      \includegraphics[width=0.45\textwidth]{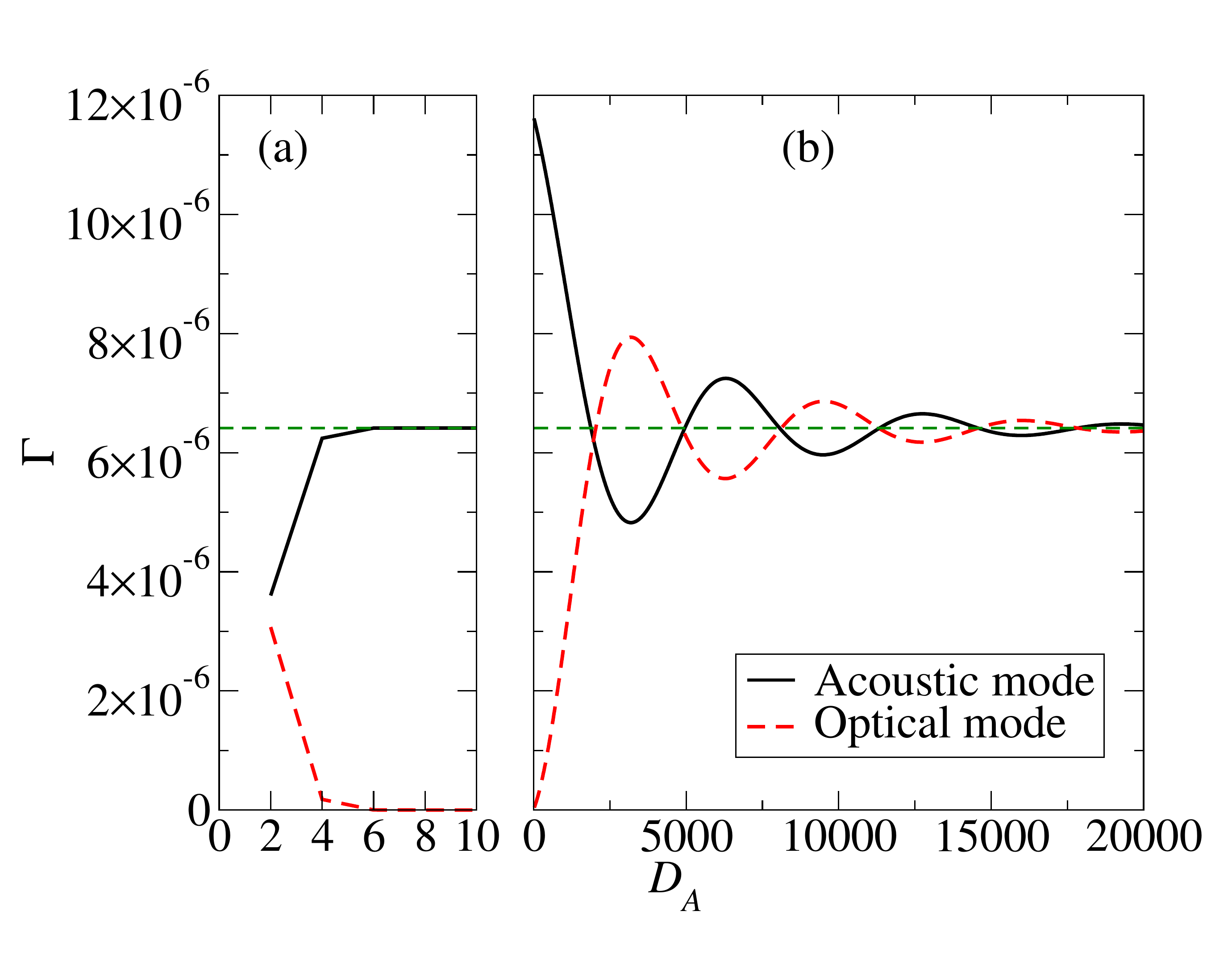}
      \caption{Full width at half maximum $\Gamma$ (in units of $t$) of the acoustic mode as a function of the horizontal distance $D_A$ between the two impurities. (a) One of the two  impurities is adsorbed to a non-VLDOS site along the line $D_Z = 1$, and the other to VLDOS sites, along the line $D_Z = 3$. The green dashed line represents the result for a single impurity adsorbed to a non-VLDOS site along the line $D_Z = 1$. We notice that the results converge very rapidly to the single impurity value in this case. (b) Both impurities are adsorbed to non-VLDOS sites along the line $D_Z = 1$. Here we observe that the convergence to the single impurity case takes place in an oscillatory way, and for very large separations only.}
      \label{FWHM}
    \end{figure}

This procedure allow us to unequivocally determine the lifetimes of the magnetic excitations in a 2-impurity system, which are characterized by the linewidths of their resonant peaks. In Fig. \ref{FWHM} we plot the linewidths of the acoustical and optical modes as a function of the impurity separation, for two distinct situations. We begin with the case in which one of the two impurities is adsorbed to VLDOS sites located along the line $D_Z = 3$, while the other sits atop the edge of the nanoribbon on a non-VLDOS site belonging to the line labeled by $D_Z = 1$. The results are depicted in Fig. \ref{FWHM}(a) and exhibit a very fast convergence to the values obtained previously for a single impurity, showing that as the separation between the impurities increases the acoustical and optical modes evolve to the ones associated with isolated impurities adsorbed to non-VLDOS and VLDOS sites. This clearly shows that when a magnetic impurity is adsorbed to a VLDOS site, its magnetic interaction with other impurities in the system becomes relatively rather weak, as expected.

The behavior changes considerably when both impurities are adsorbed to non-VLDOS sites, as displayed in Fig. \ref{FWHM}b. Here we see that as $D_A$ increases the results also converge to the value obtained for a single impurity, but in a markedly oscillatory manner and for extremely large separations only. 

The oscillations are due to quantum interferences originated in the partial confinement caused by the local potentials of two magnetic impurities. Depending on the separation between them, the presence of two impurities may either prolong or shorten the lifetime of the precession mode. The huge separation required to achieve the isolated impurity value evidences the long-range nature of the interaction between the two impurities in this case. It contrasts with the situation in which an impurity is adsorbed to VLDS sites where it becomes practically imperceptible to the other impurities in the system.

\begin{figure}
      \centering
      
\includegraphics[width=0.5\textwidth]{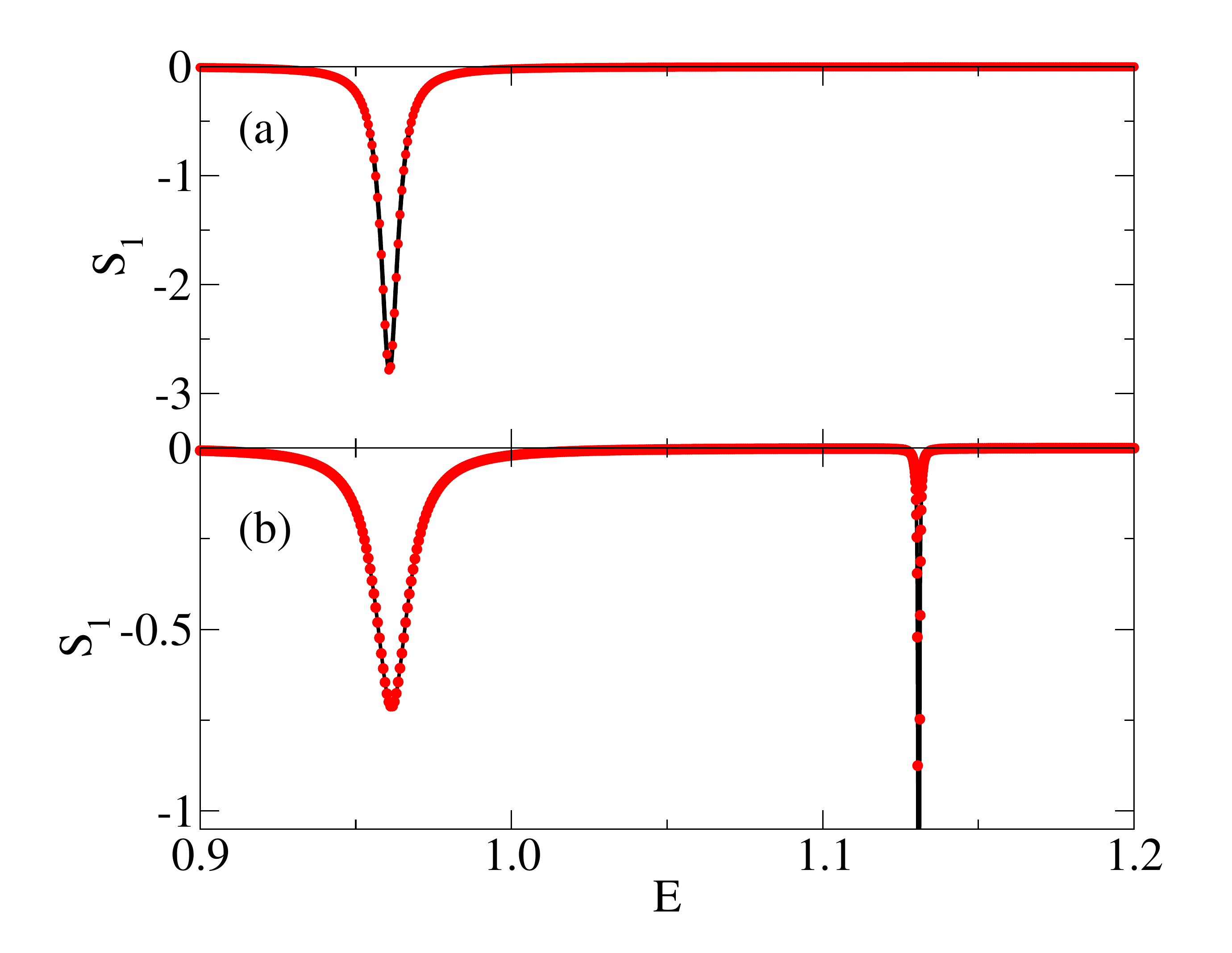}
      \caption{A comparison of the spectral lines calculated for magnetic impurities adsorbed to different lattice sites. (a) Spectral line for a single impurity adsorbed to a non-VLDOS site along $D_Z = 1$ (black line), compared with the spectral line, computed on the same site, for a system with two impurities (red dots). Here the first impurity is adsorbed to the same non-VLDOS site ($D_{Z1} = 1$), while the second one lies atop a VLDOS site at $D_{Z2}=3$, and $D_A=10$. The nanoribbon width corresponds to $N=5$. (b) Spectral line for a system with two magnetic impurities adsorbed to non-VLDOS sites labeled by $D_{Z1}=D_{Z2}=1$, and $D_A=20$, in a nanoribbon with width $N=5$ (black line). The results are compared with the spectral line calculated for a system with three impurities. Two of them are adsorbed to the same non-VLDOS sites, while the third one sits in between, atop a VLDOS site labeled by $D_{Z3}=3$, $D_{A3}=8$ (red dots). Energies $E=\hbar \omega$ are in units of $10^{-3}t$.}
    
      \label{VComparison}
    \end{figure}

Further evidence of this lack of interaction is presented in Fig. \ref{VComparison}. In panel \ref{VComparison}(a) we compare the local spectral function calculated for a single magnetic impurity adsorbed to a non-VLDOS site with the one, computed on the same site, but in the presence of a second impurity adsorbed to a VLDOS site, separated from the first by a relatively short distance. The two spectral-function lines are indistinguishable, showing that the effect of the magnetic impurity siting atop a VLDOS site on the other one is negligible. In Fig. \ref{VComparison}(b) we examine the effect of a third magnetic impurity adsorbed to a VLDOS site on the spectral line calculated for two impurities placed atop non-VLDOS sites along the line $D_Z = 1$ separated by $D_A = 20$. The third magnetic impurity is placed atop a site along the line $D_Z = 3$, at a horizontal distance $D_{A1} = 8$ from the first impurity, and $D_{A2} = 12$ from the second. Once again, the influence of the magnetic impurity adsorbed to a VLDOS site is totally negligible, showing that it is practically ``invisible'' to other impurities in the system. These results are ubiquitous for all cases examined. 

\begin{figure}
      \centering
      \includegraphics[width=0.5\textwidth]{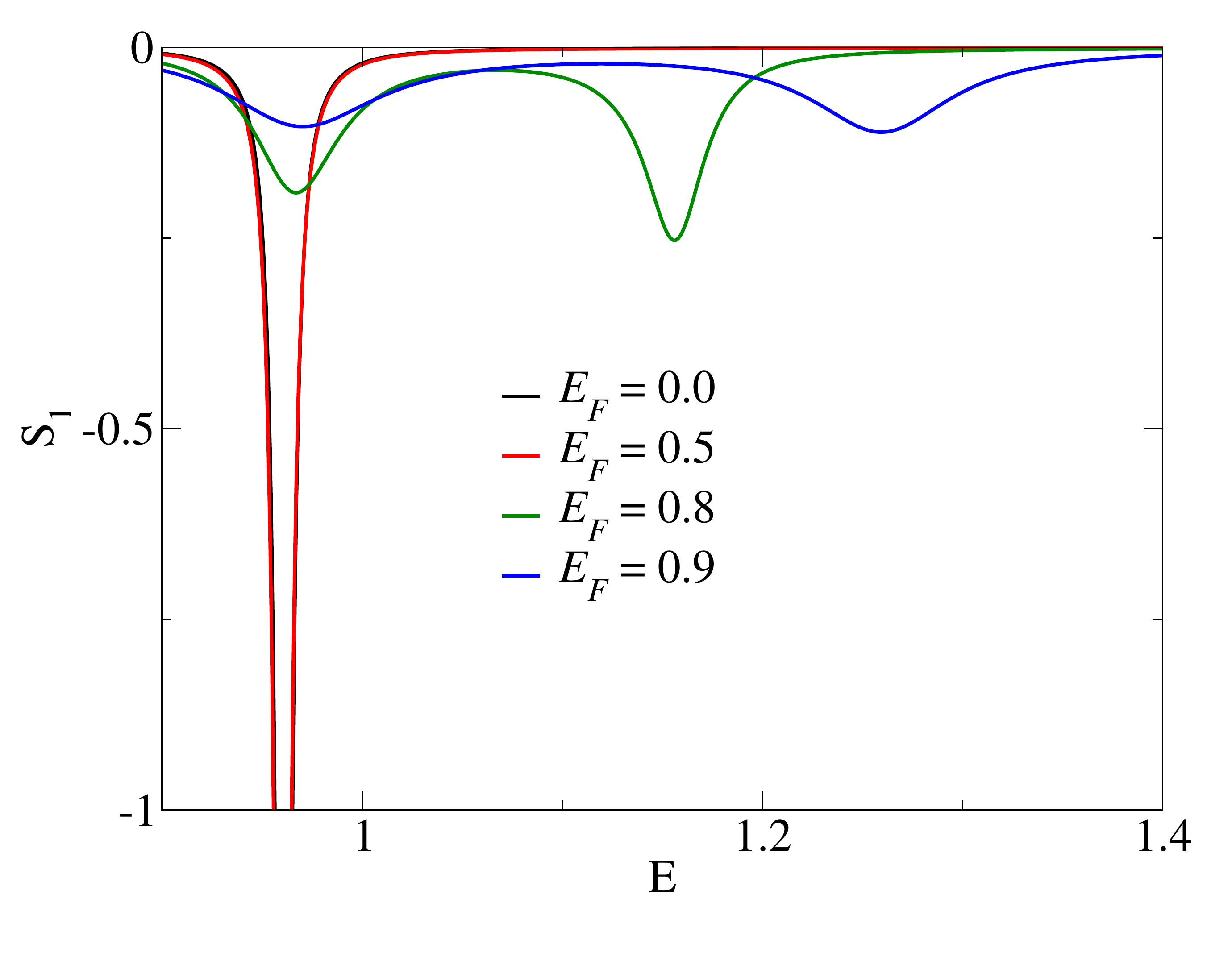}
      \caption{Imaginary part of the local susceptibility as functions of the energy $E=\hbar \omega$ (in units of $10^{-3}t$) for a system with two impurities (one non-VLDOS and VLDOS) on the non-VLDOS. $D_{Z1}=1$, $D_{Z2}=3$, and $D_A=14$, respectively, in a nanoribbon of width  $N=5$}
      \label{X2Ef}
    \end{figure}

Finally, we show that the interaction of magnetic impurities adsorbed to VLDOS sites in carbon nanoribbons with other magnetic impurities can be activated by varying the system's Fermi level. In Fig. \ref{X2Ef} we plot the local spectral functions for two magnetic impurities, one adsorbed to a non-VLDOS site and the other to a VLDOS site, for different values of $E_F$. They are placed atop two sites characterized by $D_{Z1}=1$, $D_{Z2}=3$, and $D_A=14$, respectively, in a nanoribbon of width  $N=5$. The local spectral functions are calculated at impurity adsorbed to the edge of the nanoribbon ($D_{Z1}=1$). It is clear that for relatively small values of $E_F$ the calculated spectral functions coincide with the one associated with an isolated impurity. However, once the energy gap of the VLDOS is overcome, the appearance of a ferromagnetic interaction between the two impurities becomes evident, characterized the double-peak structure corresponding to the acoustic and optical modes.  

  \section{Conclusions}
 
  In summary, we have explored the dynamical interaction between magnetic impurities adsorbed to graphene nanoribbons. We show that depending on the position the impurities are absorbed to, the interaction between them can change quite drastically. Finite-size effects in nanoribbons lead to lines of vanishing LDOS which occur in a triadic pattern from the nanoribbon's edges. We have demonstrated that magnetic impurities adsorbed to carbon atoms along such lines exhibit extremely long spin-excitation lifetimes, interacting very weakly with other impurities, and making them practically imperceptible to their counterparts. Although impurities on these lines are invisible to the interaction, we show that this can be modified by a gate voltage or through doping. In the case of magnetic impurities adsorbed to finite-LDOS sites, the spin-excitation lifetimes are considerably reduced since the angular momentum of the precessing spins can be transported away by the conduction electrons of graphene in the form of a pure spin current. The coexistence of VLDOS and non-VLDOS sites in any nanoribbon suggests that the spin excitation of ribbons doped with a finite concentration of magnetic impurities will be a mixture of these strikingly opposite responses, paving the way to controlling the spin excitation of these systems by varying the impurity concentration.

Regarding the experimental feasibility of our predictions, ferromagnetic resonance measurements are the most obvious techniques to assess the spin-excitation of our system, although it may not have enough sensitivity in the case of very dilute concentrations. However, recent experimental developments have shown that it is possible to set the magnetic moment of a nanostructure into precession by means of inelastic electron tunneling using spin-polarized STM tips \cite{Heinrich466, PhysRevLett.106.037205}. Similar method may be also employed to observe such excitations from the current modulation caused by a coherent spin precession. Furthermore, graphene is renowned for displaying large spin-diffusion lengths that can reach the $100 \mu$m range \cite{Dlubak:2012ib}. We may thus envisage the possibility of using dual-STM probe setups to investigate non-local spin excitations on a surface in the very near future \cite{PhysRevLett.112.096801,PhysRevB.90.035440,doi:10.1021/nl3042799}. To employ one spin-polarized STM tip to set a local magnetic moment into precession, while the other probes the induced excitation on another moment is challenging but experimentally feasible and seems to be the most promising method to test our claims.   

    \bibliography{biblio.bib}

\end{document}